\documentclass{article}

\usepackage{amssymb}
\usepackage{amsopn}
\usepackage{amsmath}
\usepackage{amsthm}          

\newtheorem*{theorem}{Theorem}

\title{Positivity in Rieffel's strict deformation
  quantization\footnote{Invited lecture at the XVIth International
    Congress on Mathematical Physics in Prague, August 3-8, 2009.}  
}

\author{\textbf{Stefan Waldmann} \\
  Faculty for Mathematics and Physics \\
  Physics Institute \\
  Albert-Ludwigs-University Freiburg\\
  Germany \\
  email: \texttt{Stefan.Waldmann@physik.uni-freiburg.de}
}

\date{September 2009}

\begin{document}

\maketitle

\begin{abstract}
    We review a recent result on Rieffel's deformation quantization by
    actions of $\mathbb{R}^d$: it is shown that for every state
    $\omega_0$ of the undeformed $C^*$-algebra $\mathcal{A}_0$ there
    is a continuous section of states $\omega(\hbar)$ through
    $\omega_0$.  We outline the physical interpretation in terms of
    quantization.
\end{abstract}

\section{Introduction: Continuous Fields of $C^*$-Algebras}
\label{sec:Introduction}

The aim of this short note is to recall some recently found positivity
properties \cite{kaschek.neumaier.waldmann:2009a} of Rieffel's
deformation quantization by actions of $\mathbb{R}^d$.

Quantization can be formulated in many different settings. Here, we
will focus on the following framework: starting with a $C^*$-algebra
$\mathcal{A}$, which we view as the observable algebra of the
classical system, we are interested in deformations of $\mathcal{A}$
depending on a quantization parameter $\hbar$. Though for quantization
purposes $\mathcal{A}$ is commutative, this will not be necessary in
the following. Moreover, in typical situations one has a Poisson
bracket defined on a (dense) subalgebra of $\mathcal{A}$ which gives
the ``direction of deformation''.

To make sense out of the notion ``deformation'' in this
$C^*$-algebraic framework we use continuous fields
\cite{dixmier:1977b} of $C^*$-algebras to formulate things
properly. We recall the definition: let $T$ be a Hausdorff space which
in our case will be the deformation parameter space. Suppose for every
$\hbar \in T$ we have a $C^*$-algebra $\mathcal{A}(\hbar)$. The
product $\prod_{\hbar \in T} \mathcal{A}(\hbar)$ is then equipped with
the structure of a $^*$-algebra by pointwise operations in $\hbar$.
Elements in $\prod_{\hbar \in T} \mathcal{A}(\hbar)$ will be called
\emph{sections} and are written as maps $a: \hbar \mapsto a(\hbar)$.
A \emph{continuous field structure} for $\{\mathcal{A}(\hbar)\}_{\hbar
  \in T}$ is now a collection of sections $\Gamma \subseteq
\prod_{\hbar \in T} \mathcal{A}(\hbar)$ satisfying the following
conditions:
\begin{enumerate}
\item $\Gamma$ is a sub-vector space.
\item $\Gamma$ is closed under pointwise products and
    $^*$-involution.
\item $\Gamma|_\hbar = \{a(\hbar) \; | \; a \in \Gamma\}
    \subseteq \mathcal{A}(\hbar)$ is dense for all $\hbar$.
\item \label{item:Continuity} $\hbar \mapsto
    \left|a(\hbar)\right|_\hbar$ is continuous for all $a \in \Gamma$
\item \label{item:Completeness} If $\tilde{a}$ is an arbitrary
    section which can locally in $T$ be approximated uniformly by
    sections in $\Gamma$, i.e.
    for all $\hbar_0 \in T$ and all $\epsilon > 0$
    there is a section $a \in \Gamma$ and an open neighbourhood $U
    \subseteq T$ of $\hbar_0$
    such that for all $\hbar \in U$
    \[
    \left|\tilde{a}(\hbar) - a(\hbar)\right|_\hbar \le \epsilon,
    \]
    then $\tilde{a} \in \Gamma$.
\end{enumerate}
The idea is that we axiomatically describe how ``continuous sections''
should behave. For a given $\Gamma$, the sections $a \in \Gamma$ are
called the continuous section. Note that we have \emph{not} specified
a topology on the total space $\prod_{\hbar \in T} \mathcal{A}(\hbar)$
which would be more involved. Note also, that the ``fibers''
$\mathcal{A}(\hbar)$ will in general not be isomorphic for different
$\hbar$. In particular, a continuous field is, in general, not locally
trivial and hence not a ``bundle'' of $C^*$-algebras.

The idea for quantization is now that $\mathcal{A}(0)$ is the
classical observable algebra and $\mathcal{A}(\hbar)$ for $\hbar \ne
0$ are the quantized ones. In this case, $T = [0, +\infty)$ will be
just a interval (or a suitable subset with $0$ as accumulation point).
This point of view for quantization is developed in detail by
Landsman \cite{landsman:1998a} or Rieffel \cite{rieffel:1993a}. The more
algebraic version of this deformation program with formal dependence
in $\hbar$ originates from Bayen et.  al. \cite{bayen.et.al:1978a},
see also the textbook \cite{waldmann:2007a} for a gentle introduction
and more references.

%
%

\section{Quantization of States}
\label{sec:QuantizationOfStates}

Having a deformation of the observables it is very natural to ask how
the states behave under the deformation. Here the following definition
turns out to be appropriate. Given a continuous field structure
$\Gamma$ for a collection $\{\mathcal{A}(\hbar)\}_{\hbar \in T}$ of
$C^*$-algebras we say that a collection of states
$\{\omega(\hbar)\}_{\hbar \in T}$ is a \emph{continuous section of
  states} if the map $\hbar \mapsto \omega(\hbar) (a(\hbar))$ is
continuous for all continuous sections $a \in \Gamma$, see
\cite[Def.~1.3.1]{landsman:1998a}.

We are now interested in the following situation. Let $\hbar_0 \in T$
be fixed (the classical limit) and let $\omega_{\hbar_0}:
\mathcal{A}(\hbar_0) \longrightarrow \mathbb{C}$ be a given state of
the (classical limit) $C^*$-algebra $\mathcal{A}(\hbar_0)$. If for any
given $\omega_{\hbar_0}$ it is possible to find a continuous section
$\omega(\hbar)$ of states with $\omega(\hbar_0) = \omega_{\hbar_0}$
then we call the continuous field $\Gamma$ \emph{positive}.

The physical interpretation is the following: for a positive
continuous field (which we view as a quantization of $\mathcal{A}(0)$)
every classical state is the classical limit of quantum states.
Clearly, this is very much desirable from a physical point of view as
quantum theory is believed to be the more fundamental description of
nature and hence should contain the classical description as
appropriate limit, both for the observables and the states.

Note that in general the continuous section $\omega(\hbar)$ deforming
$\omega_{\hbar_0}$ is far from being unique. There will be many
quantum states yielding the same classical limit.

%
%

\section{Rieffel's construction by actions of $\mathbb{R}^d$}
\label{sec:RieffelConstruction}

Rieffel's construction \cite{rieffel:1993a} of a deformation of
$C^*$-algebras by actions of $\mathbb{R}^d$ will give us particular
continuous fields which are nevertheless omnipresent in quantization
theory. In particular, canonical quantization of the classical phase
space $\mathbb{R}^{2n}$ can be viewed as a Rieffel deformation. We
outline the basic ideas of his construction.

Suppose a $C^*$-algebra $\mathcal{A} = \mathcal{A}(0)$ is endowed with
a strongly continuous action $\alpha$ of $(\mathbb{R}^{2n}, +)$ by
$^*$-automorphisms. Then consider the smooth vectors
$\mathcal{A}^\infty$ of this action, i.e. those elements $a \in
\mathcal{A}$ where $u \mapsto \alpha_u(a)$ is smooth. It is well-known
that $\mathcal{A}^\infty$ is a dense $^*$-subalgebra endowed with a
(finer) Fr\'echet topology. Next, choose a symplectic form $\theta$ on
$\mathbb{R}^{2n}$ and let $\hbar > 0$. For $a, b \in
\mathcal{A}^\infty$ the integral
\begin{equation}
    \label{eq:Oscillatory}
    a \star_\hbar b
    =
    \frac{1}{(\pi\hbar)^{2n}} \int
    \alpha_u (a) \alpha_v(b)
    \; 
    \mathrm{e}^{\frac{2\mathrm{i}}{\hbar}\theta(u, v)}
    \;
    \mathrm{d} u \mathrm{d} v
\end{equation}
is well-defined as an oscillatory integral and endows
$\mathcal{A}^\infty$ with a new associative product turning
$\mathcal{A}^\infty$ into a Fr\'echet algebra. Moreover, the old
$^*$-involution is still a $^*$-involution for $\star_\hbar$.  Note
that the integrand of \eqref{eq:Oscillatory} has \emph{constant} norm
since $\alpha_u$ and $\alpha_v$ are \emph{isometric}. Thus a naive
definition of the integral is not possible. One needs a more
sophisticated oscillatory integral here.

The star product $\star_\hbar$ allows for an asymptotic expansion with
respect to the $\mathcal{A}^\infty$-topology, explicitly given by
\begin{equation}
    \label{eq:Asymptotic}
    a \star_\hbar b
    \stackrel{\hbar \rightarrow 0^+}{\longrightarrow}
    \mu \circ 
    \mathrm{e}^{
      \frac{\mathrm{i}\hbar}{2}
      \theta^{kl} \partial_k \otimes \partial_l
    }
    (a \otimes b),
\end{equation}
where the partial derivatives are defined by means of the action as
$\partial_k a = \frac{\mathrm{d}}{\mathrm{d} t}|_{t=0} \alpha_{t
  e_k}(a)$ and $\mu (a \otimes b) = ab$ denotes the undeformed
product. Here $\theta^{kl}$ are the components of the Poisson tensor
associated to the symplectic two-form $\theta$. We conclude that
asymptotically, $\star_\hbar$ is the usual (formal) Weyl-Moyal star
product. In case $\mathcal{A}$ is commutative, the first order
commutator is just the canonical Poisson bracket determined by the
partial derivatives and $\theta$.

In a last step, Rieffel constructs also a $C^*$-norm
$\left|\cdot\right|_\hbar$ for all the products $\star_\hbar$ turning
$\mathcal{A}^\infty$ into a pre-$C^*$-algebra for all $\hbar$. Its
completion to a $C^*$-algebra will be denoted by $\mathcal{A}(\hbar)$.
It depends typically in a highly non-trivial way on $\hbar$. Since all
the $\mathcal{A}(\hbar)$ contain $\mathcal{A}^\infty$ as a dense
subspace it makes sense to speak of \emph{constant} sections of this
field of $C^*$-algebras. Then one shows that the constant sections
determine a continuous field structure $\Gamma$ by ``completing'' them
with respect to the requirement \ref{item:Completeness} in the
definition of a continuous field structure. This finally gives
Rieffel's continuous field.

%
%

\section{Positivity}
\label{sec:Positivity}

We are now in the position to formulate the main result
\cite{kaschek.neumaier.waldmann:2009a}: Rieffel's field is positive.
Moreover, there is even a very explicit construction of the continuous
section $\omega(\hbar)$ of states passing through a given state
$\omega_0$ of $\mathcal{A}(0)$.

We consider a fixed positive inner product $g$ on $\mathbb{R}^{2n}$
which is \emph{compatible} with $\theta$, i.e. $g(u, v) = \theta(u,
Jv)$ with a linear \emph{complex structure} $J$. Recall that a complex
structure $J$ is an endomorphism $J \in \mathrm{End}(\mathbb{R}^{2n})$
with $J^2 = - \mathrm{id}$. Note also, that such $g$ always exist.
Fixing one $g$, we consider the following convolution operator with the
Gaussian determined by $g$. For $a \in \mathcal{A}$ we define
\begin{equation}
    \label{eq:Convolution}
    S_\hbar(a)
    =
    \frac{1}{(\pi\hbar)^n}
    \int \mathrm{e}^{-\frac{g(u,u)}{\hbar}}
    \alpha_u(a).
\end{equation}
Clearly, the above integral exists in the most naive way for all $a
\in \mathcal{A}$. In fact, $S_\hbar(a) \in \mathcal{A}^\infty$ for all
$a$.

 The operator $S_\hbar$ depends now in a very nice way on
$\hbar$. In fact, we have
\begin{equation}
    \label{eq:LimitS}
    \lim_{\hbar \to 0^+} S_\hbar(a) = a
\end{equation}
for all $a \in \mathcal{A}$ with respect to the topology of
$\mathcal{A}$ and, moreover, for $a \in \mathcal{A}^\infty$ with
respect to the topology of $\mathcal{A}^\infty$. Moreover, in the
$\mathcal{A}^\infty$ topology we have for $a \in \mathcal{A}^\infty$
\begin{equation}
    \label{eq:DiffShbar}
    \frac{\mathrm{d}}{\mathrm{d} \hbar} S_\hbar(a)
    =
    \frac{1}{4} S_\hbar (\Delta_g a),
\end{equation}
where $\Delta_g$ is the Laplacian with respect to $g$ and the partial
derivatives coming from the action. From this, we immediately obtain
the asymptotic expansion
\begin{equation}
    \label{eq:ShbarAsymptotic}
    S_\hbar (a) 
    \stackrel{\hbar \rightarrow 0^+}{\longrightarrow}
    \mathrm{e}^{\frac{1}{4} \Delta_g} a
\end{equation}
in the topology of $\mathcal{A}^\infty$. This indicates that $S_\hbar$
plays the role of an equivalence transformation between the formal
Weyl-Moyal star product and the formal Wick star product
\cite{bordemann.waldmann:1997a} determined by $g$. Since the Wick star
product is known to have nicer positivity properties than the Weyl
star product, one can use the operator $S_\hbar$ to correct the
positive functionals $\omega_0$. In the formal setting this was done
earlier both for the symplectic \cite{bursztyn.waldmann:2000a} and
Poisson case \cite{bursztyn.waldmann:2005a}. In our present framework
however, the operator $S_\hbar$ is far from being invertible. The
image is in $\mathcal{A}^\infty$ and thus $S_\hbar$ is clearly not
surjective. Thus there does not seem to be a Wick type deformation
available directly.

Nevertheless, we have the following properties. For $a \in
\mathcal{A}^\infty$ a simple computation gives
\begin{equation}
    \label{eq:SonSquares}
    S_\hbar(a^* \star_\hbar a)
    =
    \frac{1}{(\pi\hbar)^n}
    \sum_{K \ge 0}
    \frac{1}{K!} \left(\frac{2}{\hbar}\right)^{|K|}
    a_K^* a_K,
\end{equation}
where the series runs over all multiindices $K = (k_1, \ldots, k_n)$ and
\begin{equation}
    \label{eq:aK}
    a_K = \int
    (z^1)^{k_1} \mathrm{e}^{-\frac{|z^1|^2}{\hbar}}
    \cdots
    (z^n)^{k_n} \mathrm{e}^{-\frac{|z^n|^2}{\hbar}}
    \alpha_{z_1 e_1 + \cdots + z^n e_n} (a)
    \mathrm{d} z^1 \cdots \mathrm{d} z^n.
\end{equation}
Here the $z_1, \ldots, z_N$ are complex coordinates induced by $J$.
The series in \eqref{eq:SonSquares} converges in the topology of
$\mathcal{A}^\infty$ and consists of \emph{squares} of the undeformed
product. But the $\mathcal{A}^\infty$ topology is finer than the
original one. This allows to conclude that
\begin{equation}
    \label{eq:ShbarSquaresToPositive}
    S_\hbar(a^* \star_\hbar a) \in \mathcal{A}^+,
\end{equation}
hence $S_\hbar$ maps squares to positive elements. In other words,
$S_\hbar$ is a positive (and even completely positive) map from the
deformed algebra to the undeformed.

Now there are two things to be checked: first, up to now the operator
$S_\hbar$ is only defined on the pre-$C^*$-algebra
$\mathcal{A}^\infty$ and not on $\mathcal{A}(\hbar)$. Nevertheless,
one can show that $S_\hbar$ is continuous with respect to the deformed
$C^*$-norm $\left|\cdot\right|_\hbar$ and thus extends to
$\mathcal{A}(\hbar)$. Clearly, the positivity properties remain true
under this completion. We arrive at a (completely) positive operator
\begin{equation}
    \label{eq:ShbarPositive}
    S_\hbar: \mathcal{A}(\hbar) \longrightarrow \mathcal{A}.
\end{equation}
Second, one checks that the $\hbar$-dependence of $S_\hbar$ is
well-behaved. In particular, on a constant and hence continuous
section $a \in \mathcal{A}^\infty \subseteq \mathcal{A}(\hbar)$ we see
that $\hbar \mapsto S_\hbar(a)$ is continuous with respect to the
$C^*$-topology on the undeformed algebra $\mathcal{A}$.  Therefor, if
$\omega_0: \mathcal{A} \longrightarrow \mathbb{C}$ is a state, it
follows that $\hbar \mapsto \omega_0(S_\hbar(a))$ is on one hand a
continuous map. Since the constant sections ``generate'' all
continuous sections we conclude that $\hbar \mapsto
\omega_0(S_\hbar(a(\hbar)))$ is continuous for all continuous sections
$a \in \Gamma$, too. On the other hand, $\omega(\hbar) = \omega_0
\circ S_\hbar$ is a state of the deformed $C^*$-algebra
$\mathcal{A}(\hbar)$ for all $\hbar \in [0, +\infty)$. Hence we proved
the following theorem \cite{kaschek.neumaier.waldmann:2009a}:
\begin{theorem}[Kaschek, Neumaier, Waldmann]
    \label{theorem:MainTheorem}
    For every classical state $\omega_0: \mathcal{A} \longrightarrow
    \mathbb{C}$
    \begin{equation}
        \label{eq:DeformedState}
        \omega(\hbar) = \omega_0 \circ S_\hbar:
        \mathcal{A}(\hbar) \longrightarrow \mathbb{C}
    \end{equation}
    defines a continuous section of states with $\omega(0) =
    \omega_0$.
\end{theorem}

There are several applications beyond the obvious ones in quantization
theory. In particular, these deformed states have been used to explore
the behaviour of the causal structure in non-commutative space-times
\cite{bahns.waldmann:2007a, heller.neumaier.waldmann:2007a}.


\end{document}